Revised

# Quantum trajectory perspective of atom-field interaction in attosecond time scale

Ivan P. Christov

Physics Department, University of Sofia, 5 J. Bourchier Str, 1164-Sofia, Bulgaria
(Fax: +3592/ 688 813, E-mail: ipc@phys.uni-sofia.bg)

**Abstract**

Here the ionization and high harmonic generation in Hydrogen and Helium by using quantum (hydrodynamic) trajectories is analyzed theoretically. The quantum trajectories are introduced in the frames of a new time dependent quantum Monte Carlo technique that allows a self-contained *ab initio* treatment of the electron correlation effects without introducing parametrized correlation potentials into the Schrödinger equation. Our approach predicts correct ionization, high harmonic spectra, and attosecond pulses generated by Helium atom beyond the single active electron approximation. It can be used to study complex multi-electron systems in arbitrary dimensions and their interaction with laser fields of both low and high intensity.

**PACS:** 42.65.Re; 42.60.Fc

## 1. Introduction

The recent progress in laser technology has led to the generation of electromagnetic pulses with duration of few femtoseconds in visible and below one femtosecond in the XUV (for recent review see [1]). The availability of such pulses with excellent space and time coherence can facilitate the time resolved study of processes which involve detailed electron motion within atoms, molecules, and nanostructures, providing in this way a wealth of new information in different fields of science and technology. The theoretical research conducted so far in the field of laser-atom interactions has been focused mainly on processes that involve ionization of atoms under the influence of powerful optical pulses. Different approaches have been used to describe the nonperturbative motion of the electrons and their interaction with the nucleus, where high order harmonics and attosecond pulses can be generated. These approaches include the *ab initio* solution of the Schrödinger equation in dimensions up to three, and for at most two electrons [2]. Semi-classical models have been developed for strong field regime, such as the strong field approximation (SFA) which assumes that before the ionization the core potential dominates while afterwards the laser field governs the electron dynamics [3]. Similar picture is encompassed by a three-step model [4] which uses classical electron trajectories. Although it provides a convenient description, because of its *ad hoc* elements the three step model cannot be considered to be a self-contained theory. On the other hand, systematic and technically difficult improvements would allow the SFA to better account for the Coulomb effects for low energy photons or electrons. Alternative approach that is aimed to describe the electron correlation in strong field ionization discards all quantum mechanical effects and employs classical trajectories, which can display the characteristic sequence of stages of non-sequential double ionization [5]. Both approaches, however, ignore possible bond-bond transitions that may take place for weaker laser field.

In most cases the calculations based on (semi-) classical trajectories are intended to substitute for the direct numerical solution of the Schrödinger equation which can be prohibitively time-expensive for multi-electron systems. Recently, a different approach to laser-atom interactions was proposed, which relies on quantum (hydrodynamic) trajectories that are calculated in a self-contained manner, together with the numerical solution of the Schrödinger equation [6]. Unlike the classical and semi-classical trajectories, the quantum trajectories are "quantum" in a sense that besides the classical forces they experience also specific quantum forces (see e.g. [7,8]). One of the most compelling reasons for using quantum trajectories is that they can provide a robust approach for systems with more degrees of freedom. It was shown in [6] that the problem for double ionization of 1D Helium atom in a strong laser fields can be reduced to solution of two 1D Schrödinger equations and two first order equations for the velocity fields of the two electrons, where the *e-e* correlation is included *ab initio* into the calculations. In principle, the reduction of any multi-electron problem to a set of single-electron problems while preserving the essential quantum dynamics is an important advantage since it allows a self-contained treatment of the interaction of complex multi-electron systems with laser fields of arbitrary strength, e.g. where there is no ionization and therefore the semi-classical trajectories may be inadequate. In this Letter we apply the quantum trajectories method to high harmonic generation (HHG) and attosecond pulse generation from one- and two-electron atoms.

## 2. Models and results

One major advantage of the method proposed in [6] as compared to the standard hydrodynamic formulation of quantum mechanics (see e.g. [7,8] and the references therein) is that it avoids the solution of the quantum Hamilton-Jacobi (hydrodynamic) equations which are nonlinear and contain quantum potentials. Instead, it solves a set of linearly coupled time dependent Schrödinger equations for the individual particles from an ensemble, concurrently with the equations for the trajectories of these particles. In this way, a new time dependent quantum Monte-Carlo framework is formed, which allows a self-contained treatment of the electron correlation effects without introducing parametrized correlation potentials into the equations of motion, as this is done in e.g. the time dependent density functional approximation (e.g. [9]). For one-electron systems where there is no electron correlation the quantum trajectories may help for visualization and interpretation of the results and may give new insights into the quantum dynamics. However, the quantum trajectories play a crucial role for multi-electron systems where the momentary position of each electron participates in the Schrödinger equations for the rest of the electrons so that the quantum waves and the classical particles participate into the dynamics on an equal footing.

### 2.1 Hydrogen

It has been shown previously that one of the most interesting regimes of atom-field interaction takes place for laser pulses with duration of a few optical cycles. Then, the cut-off harmonics of the spectrum merge in a broad band that corresponds to an isolated attosecond pulse [10,11]. First, we consider the quantum trajectories calculation for the interaction of 1D Hydrogen atom with a few-cycle laser pulse. For this purpose the following equations are solved numerically:

$$i\frac{\partial}{\partial t}\Psi(x,t)=\left[-\frac{1}{2}\frac{\partial^2}{\partial x^2}-\frac{1}{\sqrt{1+x^2}}+iA(t)\frac{\partial}{\partial x}\right]\Psi(x,t) \qquad (1a)$$

$$\frac{dx(t)}{dt}\equiv v(x,t)=\mathrm{Im}\left[\frac{1}{\Psi(x,t)}\frac{\partial\Psi(x,t)}{\partial x}\right]_{x=x(t)}-A(t) \qquad (1b)$$

Equation (1b) determines the time dependence of the quantum trajectory x(t) of an electron that is guided by the pilot wave Ψ(x,t) (see e.g.[7]), where A(t) is the vector potential. In this case, the coordinate x(t) is determined at each time step, after the wave function is been calculated from eq.1a. First, the ground state is calculated by propagating the initial trial wave-packet Ψ(x,t=0) in complex time until steady state in the electron energy is established. Concurently, an initial ensemble of randomly distributed coordinates x(t=0) is propagated according to eq.1b, where at steady state the RHS in eq.1b goes to zero and all particles stand still. It has been shown previously [6] that the statistical distribution of the particle ensemble after steady state is reached matches very well the ground state probability as calculated by the stationery wave function from eq.1a. For these parameters the ground state energy of -0.67 a.u. is obtained. Next, the time evolution in presence of the laser field is calculated. Since the strength of the field can be high, large portions of the wave function are pushed away from the nucleus and some of those may return at each reversal of the laser field. The electromagnetic radiation from the atom can be calculated by using the quantum expectation value of the dipole acceleration, and independently by using the classical radiation formula for the contributions coming from each trajectory x(t). It is important that these two approaches are independent in that the quantum trajectories x(t) are not directly related to any quantum averages, such like the dipole moment. However, the quantum trajectories can be considered to represent the motion of small volumes of the electron probability distribution in accordance with the hydrodynamic formulation [7]. Figure 1 shows the results from the calculation of the harmonic spectrum for 5 fs laser pulse with peak intensity 5 $10^{14}$ W/cm$^2$ (λ=800 nm) for zero carrier-envelope offset (CEO) in fig1a and for π/2 CEO in fig.1b. It can be seen that the spectrum calculated by using as many as 20 quantum trajectories (dashed line) matches very well the result from the dipole acceleration (solid lines) for both CEO's. This correspondence spans essentially the whole harmonic spectrum, which contrasts the predictions from the SFA where only the highest harmonics are well reproduced by the contributions from the semi-classical electron trajectories. The quantum trajectories are plotted with dotted lines in Fig.1c,d together with the attosecond pulses which are synthesized from the cut-off bands of the two spectra. It is seen from these plots that the emission of the attosecond pulse corresponds to the return of groups of quantum trajectories to the core. Clearly, these trajectories correspond to portions of the electron cloud which are accelerated by the laser pulse, as the picture is similar for zero and π/2 CEO. It is seen that besides the ionizing trajectories there are other outgoing trajectories which however do not produce attosecond pulses, which can be attributed to the lower energies of these trajectories or to the destructive interference of their contributions to the overall scattered field. In fact, the excellent agreement between the spectra calculated by using the time dependent wave-function and by using quantum trajectories proves that the latter are capable to account very precisely for the quantum motion within the electron quantum wave as it oscillates around the core. That motion is inevitably reflected to the amplitude and the phase of the scattered waves and hence to the high harmonic spectrum, which is sensitive to small fluctuations.

### 2.2 Helium

According to the model reported in [6], the interaction of 1D Helium atom with laser pulse is described by two coupled Schrödinger equations, for the two single-electron orbitals $\varphi_1(x_1,t)$ and $\varphi_2(x_2,t)$:

$$i\frac{\partial}{\partial t}\varphi_i^k(x_i,t)=\left[-\frac{1}{2}\frac{\partial^2}{\partial x_i^2}-\frac{2}{\sqrt{1+x_i^2}}+\frac{1}{\sqrt{1.5+[x_i-x_j^k(t)]^2}}+iA(t)\frac{\partial}{\partial x_i}\right]\varphi_i^k(x_i,t), \tag{2a}$$

where i=1,2 and the upper index ($k$) denotes the k-th individual particle from the i-th trajectory ensemble. The correlated motion of the two electrons is accounted for by the third term in RHS of eq.2a, which represents the time-dependent Coulomb potential experienced by each particle due to the presence of the other one. The time-dependent trajectories which enter that term are calculated by using the de Broglie-Bohm relation:

$$\frac{dx_i^k(t)}{dt} \equiv v_i^k(x_i,t) = \mathrm{Im}\left[\frac{1}{\Psi(x_i,x_j,t)}\frac{\partial\Psi(x_i,x_j,t)}{\partial x_i}\right]_{x_i=x_i^k(t);\,x_j=x_j^k(t)} - A(t) \tag{2b}$$

where the two-partite wave function of the Helium atom $\Psi(x_1,x_2,t)$ in eq.2b can be represented either as a simple product of the two spatial orbitals $\varphi_{1,2}(x,t)$ for distinguishable electrons, or as symmetrized (anti-symmetrized) product of these orbitals for equivalent electrons. We consider here both cases because they feature different physical situations that are of interest. First, the ground state of the atom is determined by propagating eqs.2a,b in complex time until steady state is established. Our model predicts ground state total energy of -2.3268 a.u. while the iterative solution of the Hartree-Fock equations gives -2.3234 a.u.. The direct diagonalization of the 2D Hamiltonian gives ground state total energy of -2.3266 a.u.. Therefore, a large portion of the ground state correlation energy is taken into account by eqs.2a,b (see also [6]). In order to elucidate the role of the electron-electron correlation potential in eq.2a we conducted detailed calculations of the HHG spectra and the degree of ionization for distinguishable and equivalent electrons. Figure 2 shows the results for HHG from Helium with two equivalent electrons, for 5 fs laser pulse with peak intensity 6.87 $10^{14}$ W/cm$^2$ and zero CEO. Although only 100 trajectories are used in that calculation the harmonic spectra generated by the two ensembles of particles and waves, according to eqs.2, are practically identical as it is expected for equivalent electrons (Fig.2a). It was verified that these spectra are also very close to the spectrum obtained from the solution for the "exact" 1D Helium where the two-dimensional Schrödinger equation is to be integrated (see e.g. [12]). The ionization is calculated by projecting the time dependent wavefunction on the ground state, thus eliminating the role of the grid boundaries (the grid size for this calculation is 200 a.u.). Figure 3 shows that the ionization close to the end of the laser pulse is practically identical for the exact solution and for our model. It is important to point out that the sensitivity of the ionization on the correlation potential is a good test for the validity of the model that we use. For example, if the smoothing parameter in the correlation potential in eq.2a is decreased from 1.5 to 1.2 the ionization changes from 19% (fig.3) up to 33.5%, again equally for both the exact solution and our model. It is seen from fig.2b that the trajectory dynamics and the attosecond pulses that correspond to the spectra in fig.2a are similar to the case of Hydrogen (fig.1a,c), that can be expected for Helium with equivalent electrons.

One additional advantage when using quantum trajectories is that these can be selected according to certain rule. That allows us to fine tune the parameters of the model for observation of different physical situations. A good example in this context is the HHG from Helium atom with distinguishable electrons. In this case the total wave function is written as a simple product of the two distinct orbitals but these are treated in a symmetric way, i.e. no single active electron approximation is used for the outer electron. Instead, trajectory selection is employed in order to distinguish the inner and the outer electron. In our calculation the trajectories are selected before the laser pulse, so that the ground energy for the particles of each couple differs by at least 0.05 a.u.. The results from the interaction of Helium atom with distinguishable electrons with 5 fs laser pulse are plotted in fig.4. In contrast to the case of equivalent electrons (fig.2), here the spectra due to the two electrons differ significantly. First, it is seen that both electrons generate harmonics with well defined cut-off. However, the intensity of the spectrum due to the inner electron is lower by two orders of magnitude than the spectrum due to the outer electron. Also, the cut-off for the inner electron is shifted to the higher frequencies by 11 harmonic orders, which corresponds exactly to the difference between the ionization potentials of the two electrons (0.597 a.u.). It is seen from fig.4b that the outer trajectories are much easier to ionize than the inner trajectories and that the attosecond pulse is generated by those outer trajectories which return to the core. Clearly, in this regime the results from the quantum trajectory calculations comply with those from the single active electron approximation, where only the outer electron is considered [2]. Besides the harmonic spectra, the ionizations also differ significantly for the inner and the outer electron, as it is seen in Fig.3. As expected, the ionization after the laser pulse for equivalent electrons is close to the mean value of the ionizations for distinguishable electrons.

## 3. Conclusions

The models developed here prove that quantum (hydrodynamic) trajectories can be very useful for calculation and visualization of the outcome of laser-atom interactions. The electron correlation effects are accounted for *ab initio* by using quantum trajectories. This formalism allows to unambiguously calculate the separate responses of the different electrons. The time dependent quantum Monte Carlo method considered here is naturally parallel and parallel efficiencies above 90% can be achieved, which significantly reduces the computation time for complex quantum systems.

## 4. Acknowledgment

The author gratefully acknowledges support from the National Science Fund of Bulgaria under contract WUF-02-05.

**Figure captions:**

**Figure 1.** Harmonic spectra (logarithmic scale) for 1D Hydrogen: (a) for zero CEO; (b) for $\pi/2$ CEO. Solid lines – from dipole acceleration, dashed lines – from quantum trajectories. The corresponding attosecond pulses (solid lines) and 1D quantum trajectories (dashed lines) are plotted in (c) and (d), where the vertical axes denote the distance from the core.

**Figure 2**. Harmonic spectra for 1D Helium with equivalent electrons (a). The contributions from the two electron ensembles are plotted with solid and dashed line; (b)-attosecond pulse (solid) and quantum trajectories (dashed).

**Figure 3.** Degree of ionization for equivalent and distinguishable electrons. The dashed line shows the “exact” result.

**Figure 4.** Harmonic spectra for 1D Helium with distinguishable electrons (a); (b)-attosecond pulse (solid) and electron trajectories (dashed).

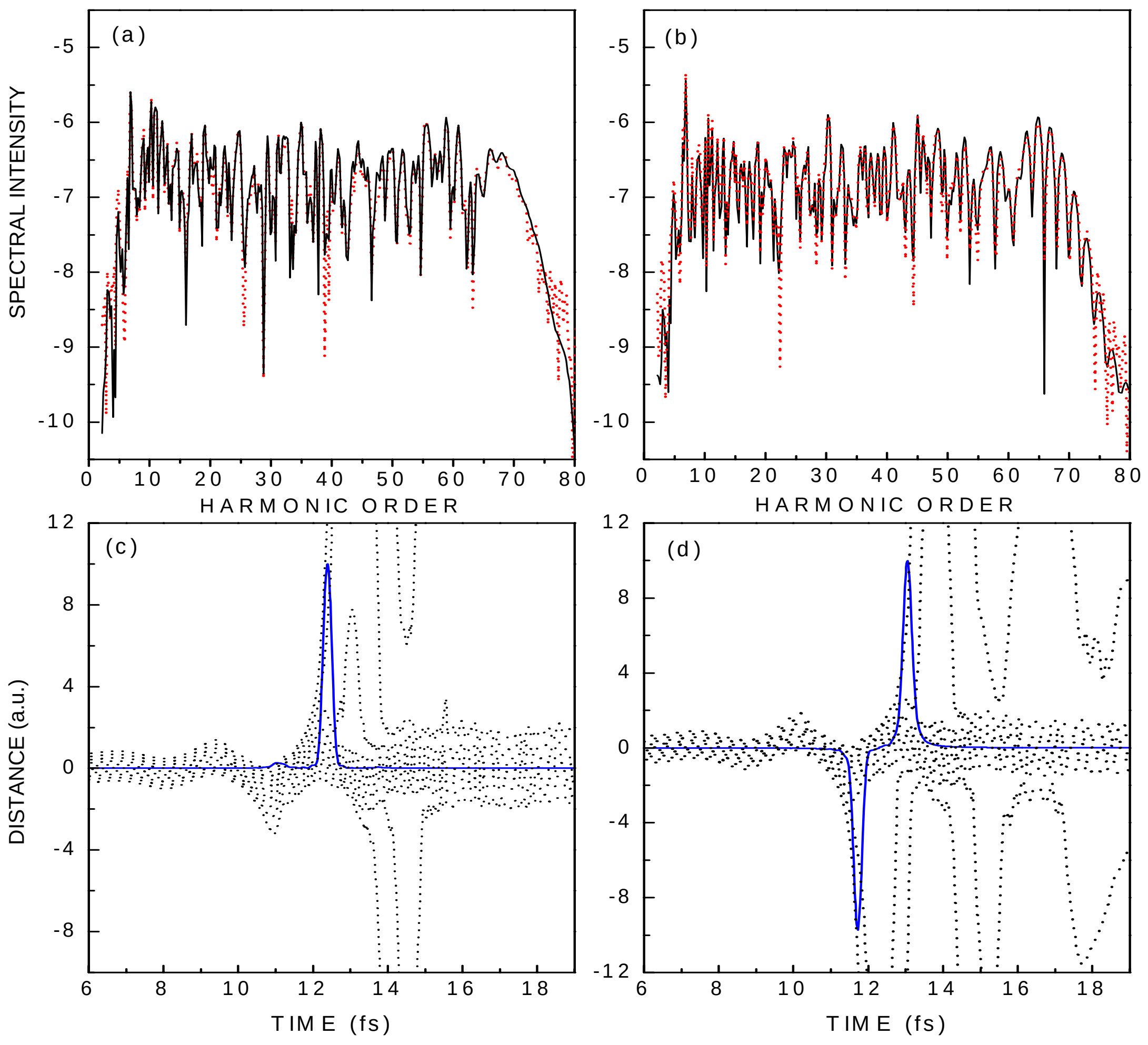

Christov, Appl. Phys. B, Fig.1

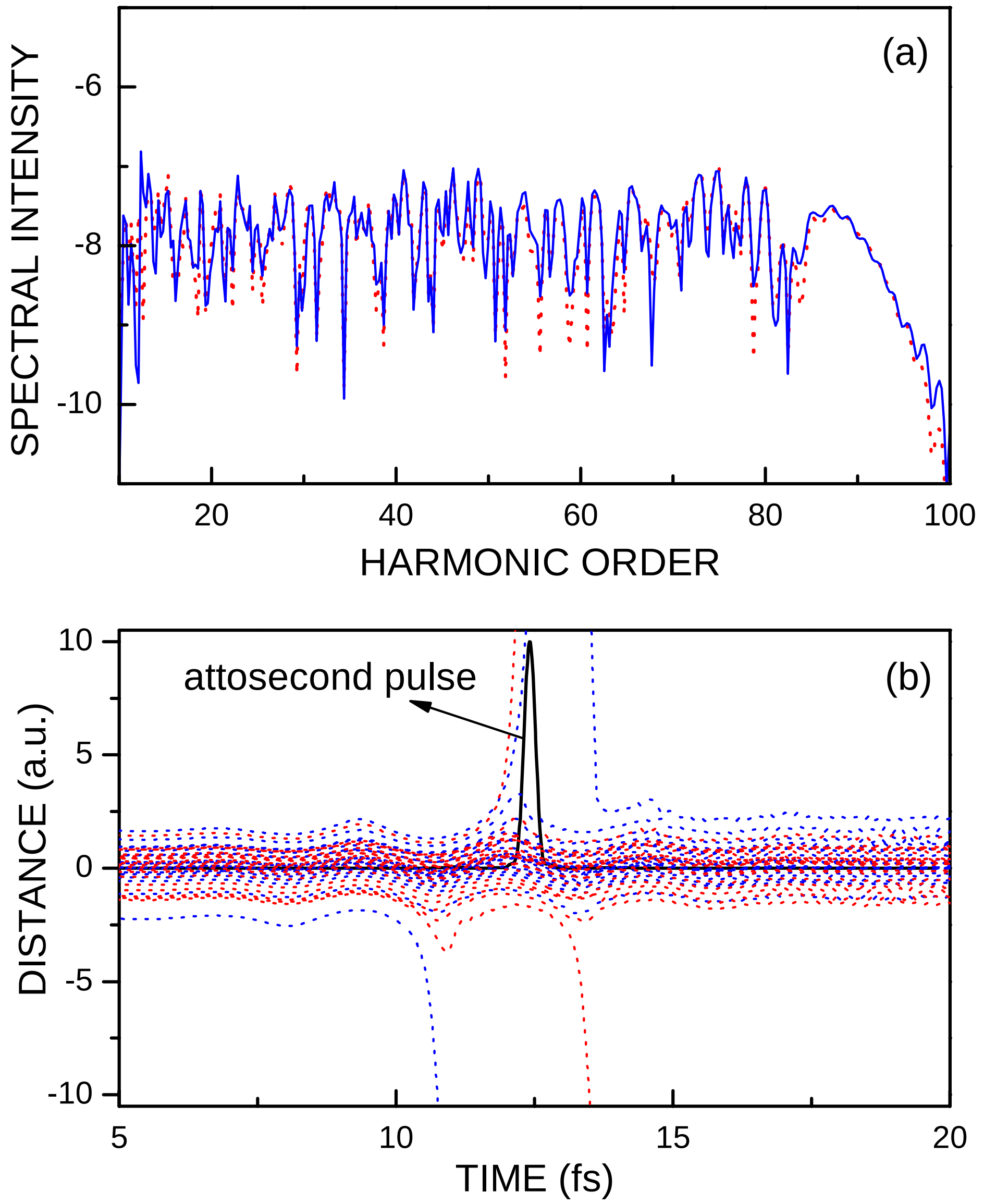

Christov, Appl. Phys. B, Fig.2

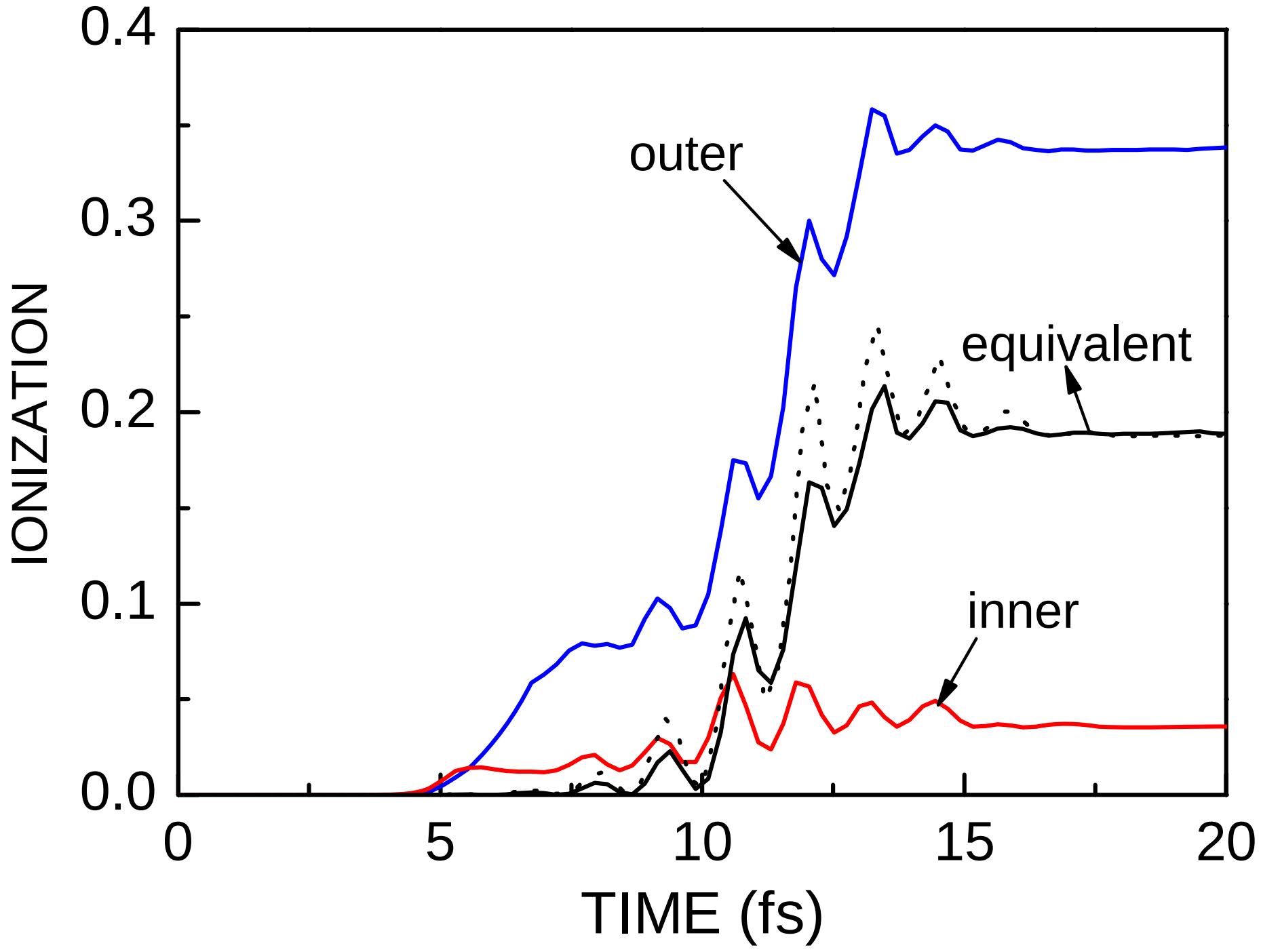

Christov, Appl. Phys. B, Fig.3

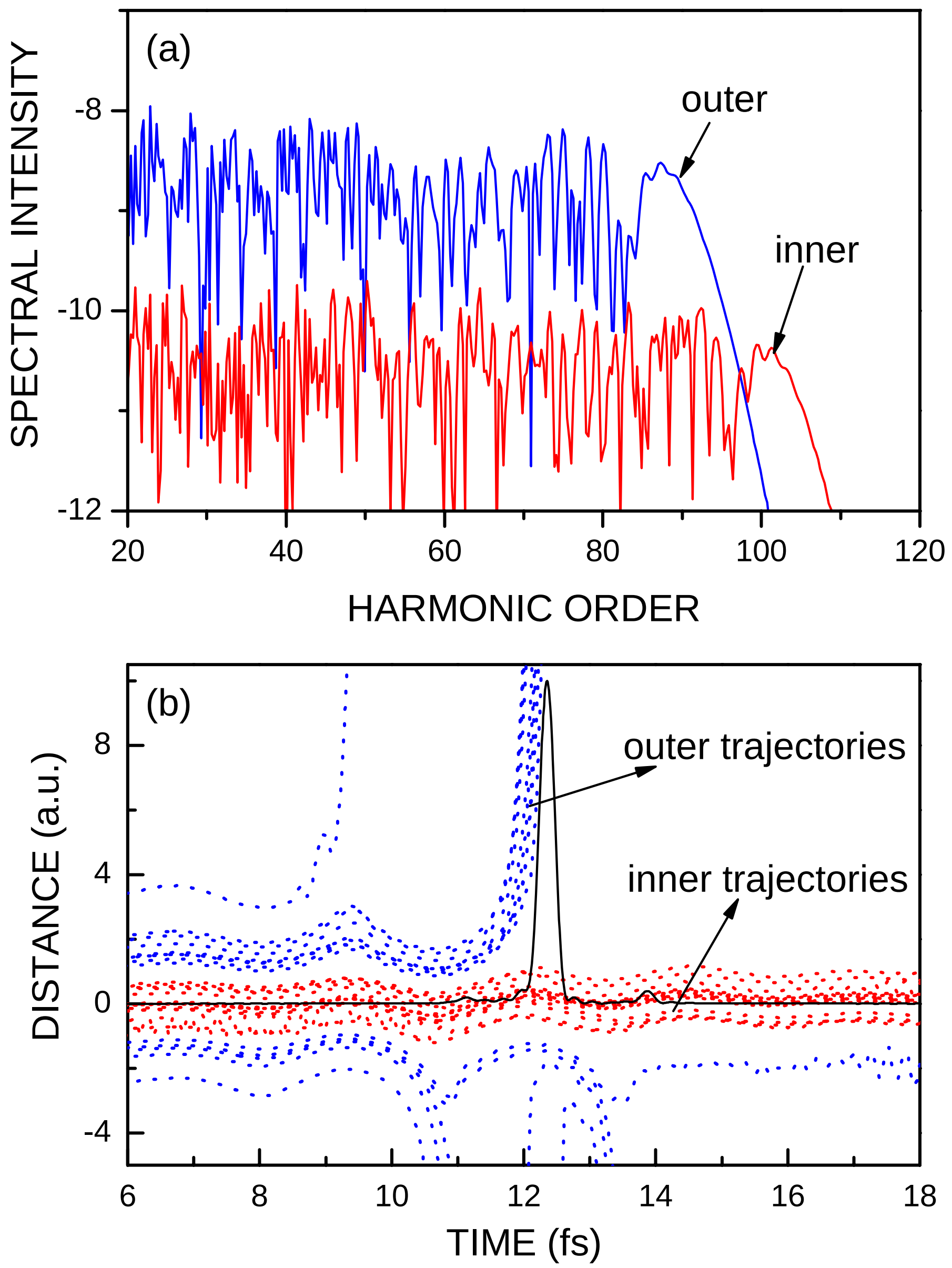

Christov, Appl. Phys. B, Fig.4